\documentclass[a4paper,11pt]{article}
\usepackage{pos}

\title{Neutrino Oscillation and Lattice QCD}

\author*[a,b]{Aaron S. Meyer}

\affiliation[a]{Department of Physics, University of California,\\
Berkeley, CA, 94720, USA}

\affiliation[b]{Nuclear Science Division, Lawrence Berkeley National Laboratory,\\
Berkeley, CA, 94720, USA}

\emailAdd{asmeyer@berkeley.edu}
\emailAdd{asmeyer.physics@gmail.com}

\abstract{
Next generation high-precision neutrino scattering experiments have the goal
 of measuring the as-of-yet unknown parameters governing neutrino oscillation.
This effort is hampered by the use of large nuclear targets: secondary interactions
 within a nucleus can confuse the interpretation of experimental data,
 leading to ambiguities about the initial neutrino interaction in scattering events.
The distribution of energies for neutrino events must instead be inferred from
 the responses of a sum of dissimilar event topologies.
For this reason, precise neutrino cross sections on nucleon targets are of
 vital importance to the neutrino oscillation experimental program.
On the other hand, the necessary experimental data for neutrino scattering
 with elementary targets are scarce because of the weak interaction cross section,
 which leads to poorly-constrained nucleon and nuclear cross sections.
Lattice QCD is uniquely positioned to provide the requisite nucleon amplitudes
 needed to enable high-precision oscillation experiments.
In particular, LQCD has the ability to probe axial matrix elements that
 are challenging to isolate or completely inaccessible to experiments.
In these proceedings, I will discuss some of my work to quantify neutrino cross sections
 with realistic uncertainty estimates, primarily focusing on
 neutrino quasielastic scattering and the nucleon axial form factor.
I will also outline how the needs of next-generation neutrino oscillation
 experimental programs can be met with modern dedicated LQCD computations.
}

\FullConference{%
  The 39th International Symposium on Lattice Field Theory (Lattice2022),\\
  8-13 August, 2022 \\
  Bonn, Germany 
}

\usepackage{xcolor}
\newcommand{\tsep}{\ensuremath{t_{\rm sep}}}
\newcommand{\chipt}{$\chi$PT}

\begin{document}
\maketitle

\section{Neutrino Oscillation}

Neutrino oscillation experiments are a major focus
 of the experimental community.
Upcoming high profile neutrino oscillation experiments,
 such as DUNE~\cite{DUNE:2020ypp} in the US
 and HyperK~\cite{Hyper-Kamiokande:2018ofw} in Japan,
 seek to measure the neutrino oscillation parameters
 to unprecedented precision.
These experiments will also measure neutrinos from astrophysical sources
 such as supernovae provide constraints on rare processes like proton decay.
To aid the experimental precision goals of these experiments,
 it is therefore important to provide support from the theory community.
One of the most promising avenues for improvement can be provided
 via constraints from Lattice QCD (LQCD).

To make predictions of neutrino event rates,
 both the neutrino flux and the neutrino cross section
 on nuclear targets are needed.
To first order, the oscillation probability for muon neutrino disappearance 
 for a given neutrino energy bin is obtained by taking the ratio of
 neutrino event rates at both the near and far detectors.
Missing muon neutrinos are converted to electron neutrinos
 that appear in the signal.
Measurements of charge-parity violation are thus obtained by comparing
 the relative neutrino to antineutrino oscillation probabilities
 for both the electron and muon flavors.

A neutrino beam is produced from decays of a secondary beam of pions.
The beam is generated by smashing a beam of protons into a fixed target,
 which produces a spray of pions at varying energy.
These pions are refocused by a magnetic horn,
 which selects pions based on their charge and energy.
There is a tradeoff between the beam intensity,
 related to the number of pions that pass the magnetic horn,
 and the width of the energy peak of the neutrino beam,
 which is dictated by how narrow the pion energy selection is.
To achieve a reasonable intensity,
 the neutrino flux must span a wide range of neutrino energies.
The neutrino energy is therefore not known before the interaction;
 even when the outgoing charge lepton is measured accurately,
 and the neutrino energy must be inferred from the byproducts of the interactions.

Additional complications arise when particles produced
 in the initial neutrino interactions rescatter before leaving the nucleus.
These intranuclear reinteractions change the particle multiplicities
 and kinematics, further obscuring the information about the initial interaction.
Even if all the particles were to escape the nucleus without reinteractions,
 neutral particles, for instance neutrons,
 can still escape the detector and go unmeasured.
For these reasons, it is not possible to know the neutrino energy
 on an event-by-event basis.
The neutrino energy must therefore be inferred from 
 neutrino event distributions under the assumption of a nuclear model.
The nuclear model, embedded in a Monte Carlo event generator,
 allows the measured distributions to be corrected back to
 an initial neutrino flux distribution as a function of energy.

For both DUNE and HyperK, the relevant energy range is so wide that
 more than one interaction process is relevant to the total neutrino cross section.
For the Monte Carlo event generators to produce
 accurate predictions of neutrino event rates,
 the cross sections for all of the relevant event
 topologies must be precisely known.
However, isolated neutrino interaction topologies are difficult to constrain
 from experimental measurements.
Neutrino interactions are either restricted to elementary targets,
 which have low statistics and are experimentally prohibitive,
 or are measured on large nuclear targets and subject to otherwise unknown
 systematic corrections from nuclear modeling.
Modern high statistics measurements on elementary targets
 are prohibitively expensive in the immediate future due to safety concerns\footnote{%
 Alternative strategies, such as subtraction of carbon cross sections from hydrocarbon data,
 are being considered.
 These strategies often rely on technical tricks
 that are not applicable to arbitrary interaction topologies.
 So far, there is no viable experimental solution for the
 lack of constraining data.}.

Alternatively, LQCD could provide constraints on matrix elements
 that are complementary to or replacements for experimental cross sections
 that difficult or impractical to measure.
First principles LQCD calculations are constructed with free nucleon targets
 and benefit from realistic uncertainty estimates that are systematically improvable.
Nucleon matrix elements obtained from LQCD are then used to constrain
 nuclear models and effective field theories,
 which allow the nucleon-level amplitudes to be extrapolated
 to the large nuclear targets used in long baseline neutrino experiments:
 water in HyperK or liquid argon in DUNE.

The nucleon axial form factor is the most popular LQCD calculation
 that is commonly applied to neutrino interactions.
This calculation gives nucleon amplitudes that are relevant for obtaining
 the nucleon charged current quasielastic (CCQE) cross section,
 the theoretically simplest and lowest-energy process that is seen
 in long baseline neutrino oscillation experiments.
CCQE is also a dominant interaction process for both HyperK and DUNE.
This amplitude is not well constrained by experiment,
 but is a relatively simple calculation for LQCD
 (in comparison to other, higher energy nucleon processes),
 making it an appealing target for LQCD computations.

These proceedings are organized as follows.
Section~\ref{sec:experiment} will discuss constraints
 coming from experimental data with nuclear targets,
 which are primarily obtained from neutrino-deuterium
 bubble chamber scattering experiments.
Section~\ref{sec:lqcd} will discuss the most recent progress from LQCD.
A detailed discussion of excited states is given in section~\ref{sec:excitedstate}.
The summary of nucleon axial form factor calculations
 is given in~\ref{sec:resultssummary},
 with some discussion of other related observables in section~\ref{sec:observables}.
The implications of results from LQCD are given in section~\ref{sec:implications}.
Some other potential calculations that will be important for neutrino physics
 are briefly discussed in section~\ref{sec:future},
 and conclusions are given in section~\ref{sec:conclusions}.

\section{Experimental Constraints on Quasielastic Scattering}
\label{sec:experiment}

The CCQE neutrino interaction converts a neutron to a proton
 by absorbing a W boson.
The weak interaction is mediated by an isovector $V-A$ interaction,
 with the matrix elements parameterized by form factors
\begin{align}
 \langle p(k+q) | V_\mu (q) | n(k) \rangle
 &= \bar{u}_p(k+q)
 \big[ \gamma_\mu F_1(q^2)
 + \frac{i}{2M_N} \sigma_{\mu\nu} q^{\nu} F_2(q^2)
 \big] u_n(k), \\
 \langle p(k+q) | A_\mu (q) | n(k) \rangle
 &= \bar{u}_p(k+q)
 \big[ \gamma_\mu \gamma_5 F_A(q^2)
 + \frac{1}{2M_N} q_{\mu} \gamma_5 F_P(q^2)
 \big] u_n(k),
\end{align}
 with 4-momentum transfer $q_\mu$, and its square $q^2=-Q^2$.
The vector current form factors, $F_1$ and $F_2$,
 are obtained from high-statistics electron scattering experiments.
For the purposes of neutrino scattering,
 the uncertainty on these form factors is assumed to be negligible.

The two form factors parameterizing the axial matrix element are
 the axial form factor, $F_A$, and the induced pseudoscalar form factor, $F_P$.
The induced pseudoscalar form factor is assumed to satisfy
 the pion pole dominance (PPD) constraint,
\begin{align}
 F_P(Q^2) = \frac{4M_N^2}{M_\pi^2+Q^2} F_A(Q^2),
\end{align}
 a relation that is empirically found to be satisfied within uncertainties.
The PPD constraint fixes the induced pseudoscalar form factor
 to be proportional to the axial form factor, leaving only the
 axial form factor as an independent function to be constrained.
The only way to probe axial matrix elements
 is via low-statistics weak interactions,
 in contrast with the high-statistics vector form factors,
 and therefore the axial form factor is the dominant contribution
 to the CCQE cross section uncertainty.
For this reason, the axial form factor is the main focus
 of studies with the goal of improving the precision
 of the CCQE cross section.

\subsection{Form Factor Parameterizations}

The parameterization of the form factor $Q^2$ dependence is
 an often underappreciated but extremely important detail
 for the neutrino oscillation community.
In experimental event rate predictions,
 the $Q^2$ dependence is either partially integrated
 into differential cross sections of directly measurable kinematic variables,
 or completely integrated away to produce total cross sections.
Though the momentum transfer $Q^2$ is convenient to theorists,
 it cannot be measured directly by neutrino scattering experiments
 and so its extraction incorporates a nonnegligible amount of model dependence.
If event rate predictions are based on model dependent cross sections,
 then that model dependence is implicitly introduced into
 the procedure of neutrino energy reconstruction.
Using constraints from one experiment to make event rate predictions for another
 then risks combining nuclear models in a way that may not be entirely self consistent.

Despite the low statistics,
 neutrino scattering data have historically been used to
 claim a 1\% uncertainty on the axial form factor~\cite{Bodek:2007ym}.
For comparison, this is more precise than the uncertainty on the vector form factors,
 which are constrained with several orders of magnitude more interaction events.
The aggressive error budget is a result of overconstraining the axial form factor
 with the restrictive dipole ansatz,
\begin{align}
 F^{\rm dipole}_A(Q^2) = g_A (1+Q^2/m_A^2)^{-2}.
\end{align}
This parameterization has only two free parameters:
 the axial vector coupling, $g_A=F_A(0)$,
 and the axial mass parameter, $m_A$.
This ansatz has remained popular largely because it reproduces
 the correct $Q^2\to\infty$ scaling behavior,
 falling off as $Q^{-4}$ at large $Q^2$ as expected from perturbative QCD.
However, this scaling behavior takes over well outside of the kinematic
 region probed by neutrino oscillation experiments.
Even if the power of the scaling is correct,
 there is no guarantee that the prefactor on the scaling in that regime is correct.
The dipole ansatz violates QCD unitarity bounds,
 meaning the dipole is inconsistent with the theory it is being used to describe.

In place of the dipole ansatz,
 the model independent $z$ expansion~\cite{Bhattacharya:2011ah}
 is an alternative used to study the form factor uncertainty.
The $z$ expansion parameterization is a conformal mapping
 of the squared momentum transfer $Q^2$,
\begin{align}
 F_A(z) &= \sum_{k=0}^\infty a_k z^k, \\
 z(Q^2) &= \frac{
 \sqrt{t_c+Q^2}-\sqrt{t_c-t_0\vphantom{q^2}}}{
 \sqrt{t_c+Q^2}+\sqrt{t_c-t_0\vphantom{q^2}}},
\end{align}
 where $t_c$ at least as small as the particle production branch cut threshold
 ($9M_\pi^2$ for the axial form factor).
The parameter $t_0$ fixes the intercept $z(Q^2=-t_0)=0$ and can be chosen for convenience,
 often chosen to minimize the value of $|z|$ below the maximum $Q^2$
 probed by the experiment.
Parameterized in this way, the kinematically allowed region of spacelike
 $Q^2_{\rm max} >Q^2>0$ is equivalent to the restriction $|z|<1$,
 guaranteeing a small parameter expansion.
QCD unitarity conditions also require that the $a_k$ be bounded and decreasing,
 so the $Q^2$ behavior of the form factor is mostly captured in the lowest
 order coefficients.
The expansion may therefore be truncated at a finite order without
 introducing large systematic effects.
Additional, less relevant fit parameters can then be introduced naturally by
 simply increasing the order of the truncation.

To meet the needs of the experimental community,
 the LQCD community must provide a complete form factor
 parameterization as a function of $Q^2$,
 including a covariance matrix for the fit parameters.
It is necessary to provide more than just an axial mass parameter
 to improve upon experimental constraints for the axial form factor.
There is no reason to expect the dipole ansatz to be an accurate
 representation of the form factor shape, and there are many reasons
 to suspect its failure even with existing imprecise constraints.
While an inflated axial mass may produce an uncertainty band
 large enough to cover the possible range of cross section values
 at any fixed $Q^2$, it also imposes strong, unphysical correlations
 between values at different $Q^2$.
Integration over incomplete ranges of $Q^2$,
 for instance when converting $Q^2$ to other kinematic variables,
 is sensitive to these correlations between different momentum transfers.
Overly restrictive form factor parameterizations would therefore
 introduce invisible and undesired bias into measurements
 of cross sections and event rates.

\subsection{Neutrino-Deuterium Scattering}

Elementary target constraints on the axial form factor
 are obtained from neutrino scattering in deuterium bubble chamber experiments%
\footnote{%
 Other constraints on the nucleon axial form factor
 have historically been obtained from electro pion production data.
 A review of these data can be found in Ref.~\cite{Bernard:2001rs}.
 These extractions are only strictly valid close to $Q^2=0$ and $M_\pi=0$.
 Considering data away from these limits introduces model dependence,
 which is larger than the statistical uncertainty and ignored in previous global fits.
 Allowing for model dependence, pion production data
 are consistent with and no more constraining than both deuterium and LQCD extractions.}.
These experiments~\cite{Baker:1981su,Miller:1982qi,Kitagaki:1983px}
 were carried out in the late 1970s to early 1980s
 and have only about $10^3$ CCQE events each.
Despite the popularity of the deuterium bubble chamber results,
 their utility is limited in contemporary applications.
The original data are lost;
 only the event distributions (without their correlations)
 obtained from digitizing the plots in the original publications remain.
There are also unknown corrections that have been applied to these
 data to account for systematic corrections.

The deuterium scattering data were reanalyzed with
 both the dipole and $z$ expansion parameterizations in Ref.~\cite{Meyer:2016oeg}.
In this work, the dipole was found to underestimate
 the axial form factor uncertainty by at least a factor of 10.
The $z$ expansion axial form factor parameterization
 produces a CCQE cross section with a more realistic 10\% uncertainty
 and a central value that lies outside of the dipole parameterization's
 1\% uncertainty band.
The significantly larger form factor uncertainty,
 when propagated to a nuclear target cross section,
 can be the same order as discrepancies between
 cross section predictions from theory and experimental measurements.
This blurs the distinction between nucleon and nuclear uncertainties;
 tensions that would have been entirely attributed to nuclear modeling
 with an unphysically small dipole form factor uncertainty
 may actually be from some unknown combination of nucleon and/or nuclear origins.

In addition to the previously stated hindrances, the corrections applied to the
 deuterium scattering data in order to account for the presence of
 a spectator proton are also lacking.
These deuterium corrections are assumed to be energy independent,
 which is an unfortunate side effect of the nuclear models
 losing validity for the relatively high energies at which these experiments operated.
The deuterium corrections that were imposed were also mild,
 becoming unity above $Q^2\gtrsim0.2~{\rm GeV}^2$.
More aggressive choices for the deuterium $Q^2$ dependence were tried,
 but did not make noticeable differences to the fits.
All of the mentioned details are potential reasons to be suspicious
 of the constraints coming from these scattering data,
 an observation that should be kept in mind when comparing to
 LQCD data in Sec.~\ref{sec:resultssummary}.

\section{LQCD Axial Form Factor}
\label{sec:lqcd}

LQCD calculations of the nucleon axial form factor have come a long way
 within even the past few years.
Simulations of the nucleon axial form factor have been carried out
 with complete error budgets including large volumes and physical pion masses.
One of the major achievements of the lattice community is the progress
 toward understanding the excited state contamination to the nucleon
 axial form factor.
This has been a multi-prong effort,
 approaching the problem both with chiral perturbation theory (\chipt)
 as well as with dedicated LQCD systematics studies.
Due to this progress,
 many of the open questions regarding consistency checks
 have been resolved, including the historically low $g_A$\footnote{%
 For recent reviews, please refer to the FLAG review~\cite{FlavourLatticeAveragingGroup:2019iem}
 or the USQCD white paper~\cite{Kronfeld:2019nfb}.}
 and the apparent violations of the partially conserved axial current (PCAC)
 and PPD relations for the nucleon form factors.

\subsection{Excited State Contamination in the Axial Form Factor}
\label{sec:excitedstate}

One of the important considerations when dealing with excited states
 is the source-sink separation time, \tsep.
Studies of the dependence on \tsep~were largely motivated by calculations of $g_A$.
Strategies for addressing excited state contamination generally fall
 into one of two categories: either
 1) relying on large \tsep~to extract the asymptotic large-time nucleon ground state,
 such as the summation method~\cite{Capitani:2012gj},
 or 2) using many \tsep~to fully parameterize the time dependence
 of the excited state contamination,
 such as the case in variational methods or multiexponential fits.
Although the former is conceptually simpler in LQCD calculations,
 \chipt~\cite{Bar:2016uoj,Bar:2017kxh} suggests that the ground state saturation
 may not be reached until at least $\tsep > 1.5~{\rm fm}$,
 and empirically even up to $\tsep > 2.0~{\rm fm}$~\cite{He:2021yvm}.
Reliance on large \tsep~therefore battles with the exponential
 signal-to-noise degradation common to nucleon correlation functions,
 making this strategy less cost effective than other methods.
In contrast, incorporation of many \tsep~allows
 for parameters to be constrained by data with
 better statistical precision.
In addition, more values of \tsep~are needed to fully assess systematics
 associated with excited states; at least three values of \tsep~are needed
 to have a 0 degree of freedom constraint on the matrix elements
 connecting the ground state to a single excited state.

The excited state contamination in both $g_A$
 and the axial form factor is now believed to be mostly
 from nucleon-pion ($N\pi$) scattering states.
This was first demonstrated in a LQCD computation in Ref.~\cite{Jang:2019vkm}.
Effects from these nuisance $N\pi$ states can be substantial.
The $N\pi$ contributions had previously gone unnoticed
 because of how they contribute to correlation functions.
These contributions are small in the two-point correlation functions
 since the three-quark interpolating operator has a relatively small overlap
 onto five-quark states that include a pion.
However, the axial current acts like a pion creation operator,
 enhancing the overlap with $N\pi$ intermediate states in the
 three-point correlation function sufficiently for them to become appreciable.
The lack of evidence for $N\pi$ states in the two-point correlators
 is therefore insufficient to rule out their presence in the three-point functions,
 a mistake that has been made in previous generations of LQCD calculations.

Although \chipt~formally suggests there is a $L^{-3}$ suppression factor
 on the overlap factors for $N\pi$ states~\cite{Bar:2017kxh},
 their overall effect can remain large with increasing volume
 because of a compensating effect from the density of states.
This is because the smallest unit of lattice momentum is $2\pi/L$,
 which decreases as the lattice volume increases.
Thus, the number of allowed $N\pi$ momenta combinations
 that both 1) remain below some desired energy threshold,
 and 2) sum to the total center of mass momentum,
 increases proportional to the volume.
Although the effects from a \emph{single} $N\pi$ state
 are subject to power law suppression with the lattice volume,
 the \emph{total} of all $N\pi$ states to those correlators
 scales as a mild exponential volume correction.

Operator relations between the correlation functions have
 been a vital tool for resolving these issues.
The PCAC relation connects the axial matrix element to the pseudoscalar matrix element,
\begin{align}
 \partial^\mu A_\mu = 2 m_q P.
\end{align}
This is an operator relationship that is exact in the continuum limit.
Sandwiching the PCAC relation with incoming and outgoing nucleon states
 yields a different connection for the form factors,
\begin{align}
 2M_N F_A(Q^2) - \frac{Q^2}{2M_N} F_P(Q^2) = 2m_q F_5(Q^2),
\end{align}
 with pseudoscalar form factor $F_5$.
This is referred to as the generalized Goldberger-Treiman (GGT) relation.
Since this relationship is sensitive to the accuracy of the form factor extraction,
 it provides a useful consistency check of systematics in
 nucleon form factor calculations.
In essence, if the $N\pi$ states are improperly subtracted away from the
 nucleon matrix elements, then (barring some coincidental cancellation)
 they will appear as a contaminant that spoils the GGT relation.

An important consideration when assessing the satisfaction of the GGT
 relation is the kinematic dependence of the $N\pi$ states.
This was studied extensively with \chipt~\cite{Bar:2018xyi,Bar:2019gfx,Bar:2019igf}.
The axial form factor receives corrections from pion loops,
 which can result in approximately a 5\% shift to the axial form factor
 at Euclidean times as large as 2~fm, essentially independent of $Q^2$.
In contrast, the induced pseudoscalar form factor receives tree level
 corrections from pions, resulting in a dramatic $Q^2$ dependence
 that can be as large as 40\% at low $Q^2$ and falling off rapidly.
The pseudoscalar contamination is obtained from the combination of the two
 others via application of the GGT relation.
The predicted $Q^2$ falloff of the induced pseudoscalar
 $N\pi$ form factors bears a striking resemblance to the
 reported apparent deviations from the GGT relation,
 giving empirical (yet insufficient) evidence
 that the $N\pi$ states may indeed be responsible for the deviation.

The question still remains: how should the $N\pi$ contamination be dealt with?
One strategy is to examine the temporal axial current $A_4$,
 which has a comparatively large coupling to the $N\pi$ states,
 and use those couplings to determine the $N\pi$ states in the spatial axial currents.
Alternatively, if the contamination from $N\pi$ states is dominated
 by contributions from the induced pseudoscalar form factor, it might be better
 to only constrain the axial form factor with correlation functions where
 the induced pseudoscalar contributions vanish
 (i.e. momentum perpendicular to the spatial axial vector direction).

The gold standard method for demonstrating the (in)significance
 of contamination from the $N\pi$ states is a calculation with
 explicit $N\pi$-like interpolating operators.
Empirical observations from meson calculations~\cite{Wilson:2015dqa}
 have shown that the only way to get appreciable
 overlap onto multiparticle states is to include interpolating operators
 with quark operators that mimic the desired particle content.
While constraints of $N\pi$ contributions may not be exact without
 these interpolating operators, their effect on the ground state signal
 might also not be relevant within the quoted uncertainties.
If the effects can be shown to be small with an explicit calculation,
 then there will be less pressure for the community to perform several suites of
 calculations including full operator bases of 3- and 5-quark interpolators.
Even if the $N\pi$ operators do prove to be necessary,
 \chipt~predicts that their effects will be strongest at low $Q^2$.
Regardless, there is value in assuming that the $N\pi$ are properly handled
 and studying the effect of state of the art LQCD calculations on
 the neutrino event rate predictions.

\subsection{Summary of Results}
\label{sec:resultssummary}

Nucleon axial form factor calculations are now appearing that have
 complete error budgets, including full chiral-continuum
 and finite volume extrapolations in addition to detailed studies
 of excited state systematics.
This enables detailed comparisons of LQCD calculations for
 consistency both against each other and against experimental
 extractions of the nucleon axial form factor.

\begin{figure}[htb!]
\centering
\includegraphics[width=0.8\textwidth]{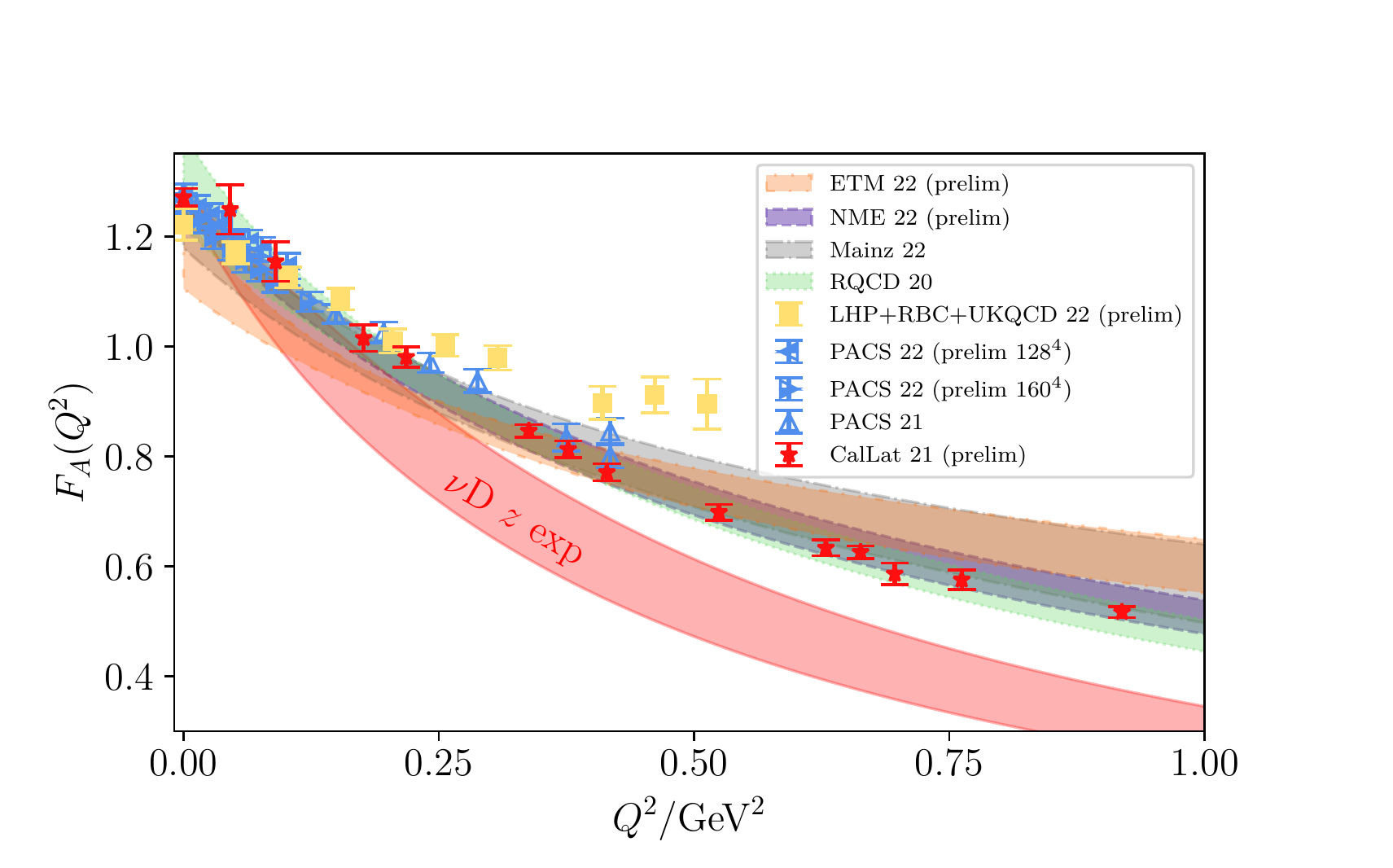}
\caption{
A summary of the most recent LQCD calculations of the nucleon axial form factor
 as a function of $Q^2$ compared against the deuterium scattering result.
The red band, labeled ``$\nu$D $z$ exp,''
 is the axial form factor result from
 deuterium bubble chamber scattering in Ref.~\cite{Meyer:2016oeg}.
The remaining bands and scatter points are all LQCD results at physical pion mass.
Scatter points indicate results obtained on a single lattice ensemble.
The error bands have a complete error budget
 including extrapolation to the continuum, chiral, and infinite volume limits.
Adapted from Ref.~\cite{Meyer:2022mix}.
\label{fig:gaq2-summary}
}
\end{figure}

The most recent existing LQCD computations at physical pion mass
 of the nucleon axial form factor as a function of $Q^2$
 are represented in Fig.~\ref{fig:gaq2-summary}.
The error bands
 (ETM 22~\cite{Alexandrou:2020okk},
 NME 22~\cite{Park:2021ypf},
 Mainz 22~\cite{RQCD:2019jai},
 RQCD 20~\cite{Djukanovic:2022wru})
 show results with a complete error budget,
 including chiral, continuum, and finite volume extrapolation
 as well as systematics to account for excited state contamination.
These results are all in good agreement with each other,
 demonstrating the consistency of the LQCD extrapolations.
The scatter points
 (LHP+RBC+UKQCD~\cite{Abramczyk:2019fnf},
 PACS 22 ($128^4$)~\cite{Tsuji:2022ric},
 PACS 22 ($160^4$)~\cite{Tsuji:2021bdp},
 PACS 21~\cite{Shintani:2018ozy,Ishikawa:2021eut},
 and CalLat 21~\cite{Meyer:2021vfq})
 denote single-ensemble results.
Although these results could have unquantified systematic biases,
 the results are all still in agreement with each other and with the fully
 extrapolated results, suggesting that the unknown systematics are likely to be small.

Fig.~\ref{fig:gaq2-summary} also shows the neutrino deuterium
 scattering result from Ref.~\cite{Meyer:2016oeg},
 labeled as ``$\nu$D $z$ exp.''
The comparison is interesting:
 the fully extrapolated LQCD result bands are in good agreement
 with the deuterium results at low $Q^2$,
 but as much as $3\sigma$ high for $Q^2 \gtrsim 0.75~{\rm GeV}^2$.
The excited state contamination from $N\pi$ states predicted by \chipt~are
 expected to be worst at low $Q^2$, where the agreement is best.
This means that the $N\pi$ contaminations,
 at least so long as \chipt~can be trusted at such large $Q^2$,
 are not responsible for the discrepancy.

There are several possibilities of effects from the LQCD calculations
 that could be responsible for this tension.
Apart from the previously mentioned $N\pi$ states and the breakdown of \chipt,
 there could be some other excited state contamination not represented
 in the \chipt~results.
Systematic uncertainties from extrapolation could be a contributing factor
 if the uncertainties from those extrapolations are underestimated.
For example, this could be due to the infinite volume extrapolation;
 the discrete lattice momenta become denser as the volume gets larger,
 so calculations must include more momenta to achieve the same
 momentum transfer $Q^2$ in physical units.
If the same number of momenta are computed
 without also selecting momenta across a range that scales with the volume,
 then the high $Q^2$ extrapolation could fall outside of the 
 fit region for larger volumes, which could spoil the extrapolation for high $Q^2$.

\subsection{Sample Observables}
\label{sec:observables}

The squared axial radius is defined by the derivative of the form factor at $Q^2=0$,
\begin{align}
 r_A^2 = -\frac{6}{g_A} \frac{dF_A}{dQ^2} \biggr|_{Q^2=0}.
\end{align}
This quantity has implications for experiments that deal with small momentum transfers,
 such as neutron decay.
Like the axial form factor, current estimates of the axial radius are imprecise
 and can be more strongly constrained by LQCD calculations.

\begin{figure}[htb!]
\includegraphics[width=0.55\textwidth]{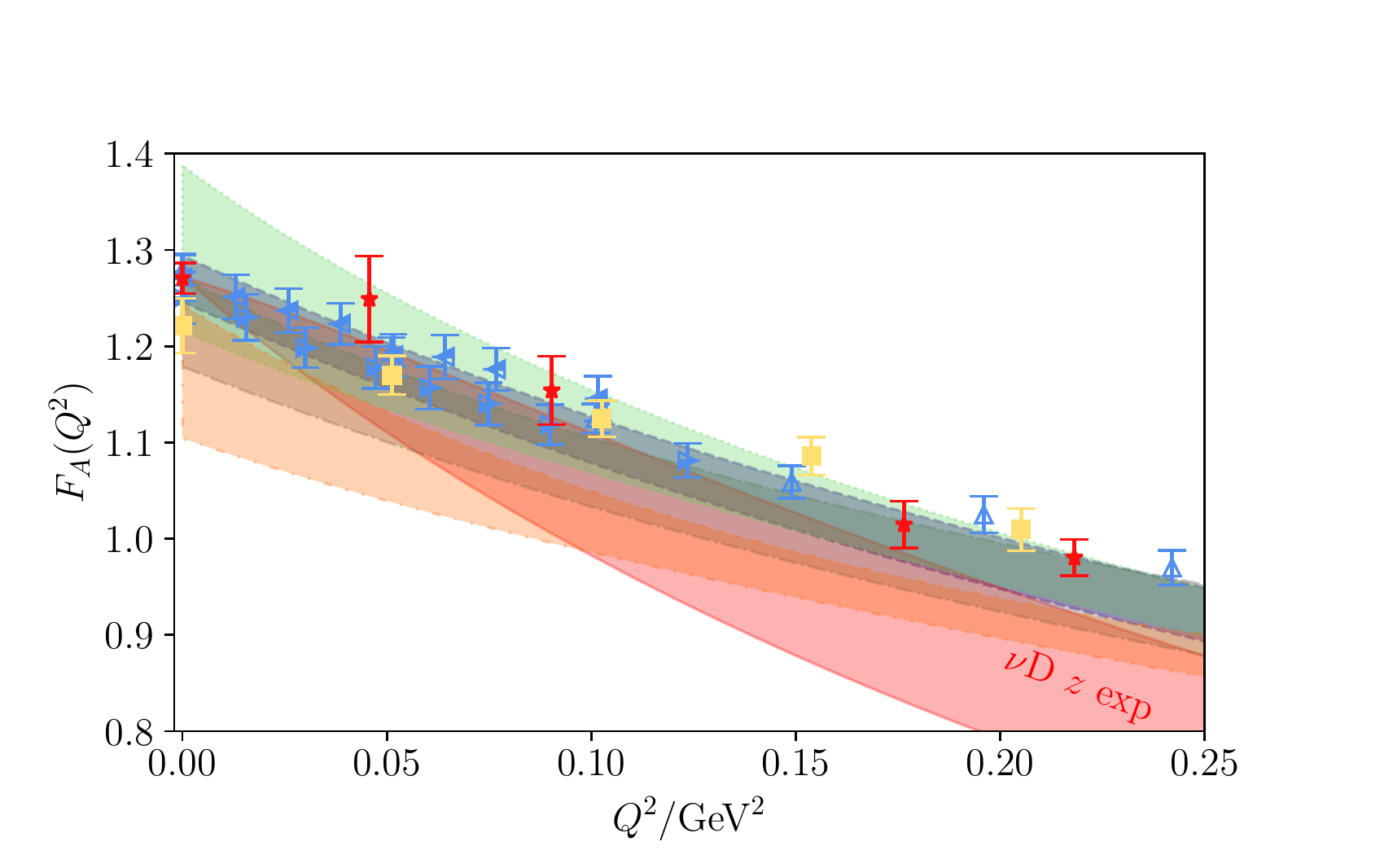}
\hspace{-4em}
\includegraphics[width=0.55\textwidth]{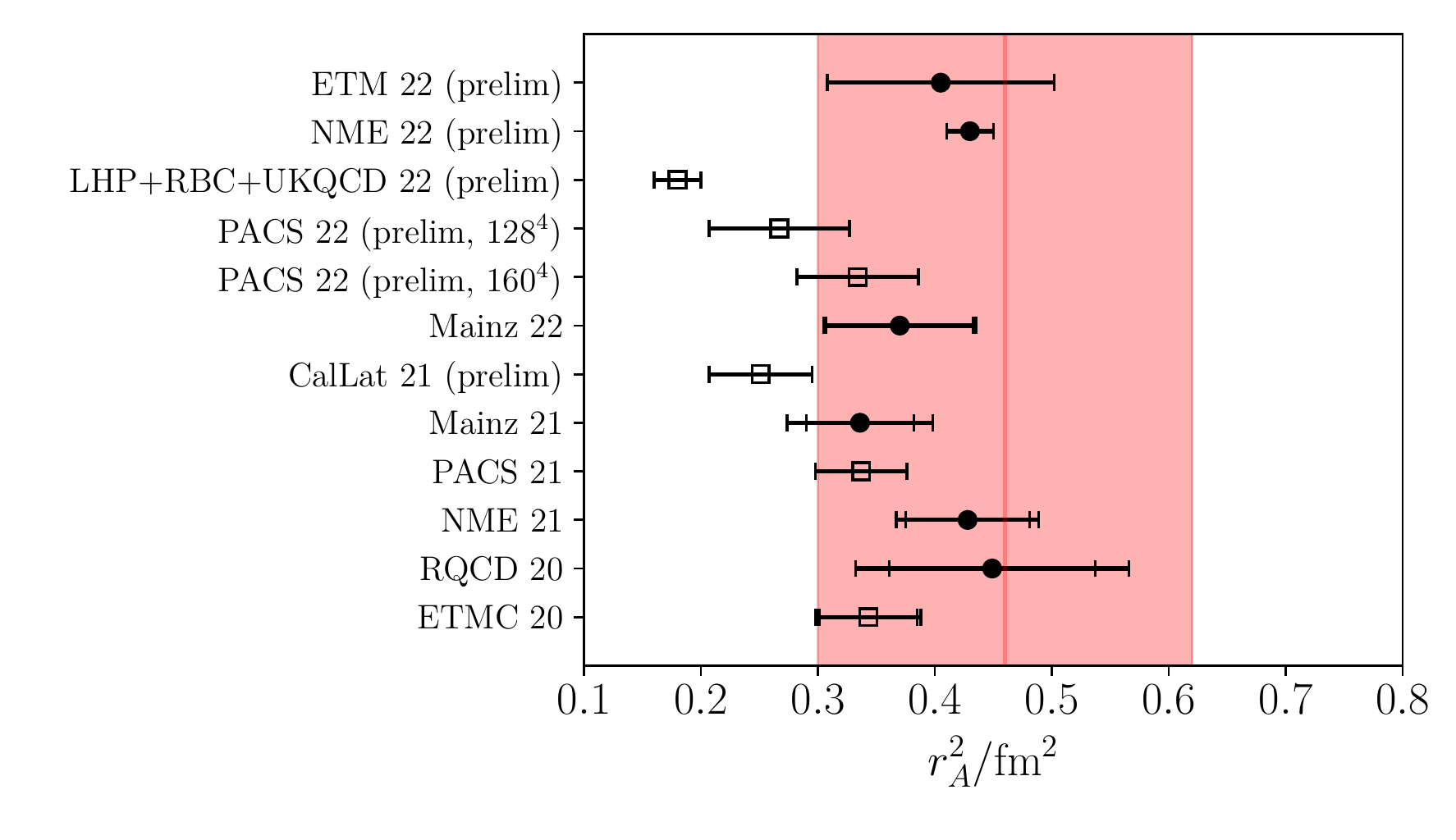}
\caption{
Left: the same as Fig.~\ref{fig:gaq2-summary}, but zoomed close to $Q^2=0$.
Right:
 a summary of the existing estimates for the squared axial radius parameter
 obtained from LQCD calculations.
Filled circles (open squares) correspond to results with (in)complete error budgets.
The red band comes from an average of deuterium scattering
 and muonic hydrogen~\cite{Hill:2017wgb}.
\label{fig:ra2}
}
\end{figure}

The current status of LQCD estimates of the squared axial radius
 are shown in Fig.~\ref{fig:ra2}.
In this plot, LQCD results with (in)complete error budgets are
 shown as filled circles (unfilled squares).
These results are compared against an average of
 the axial radius extracted from experimental settings in Ref.~\cite{Hill:2017wgb},
 including both deuterium scattering and muonic hydrogen.
Although the uncertainties are large,
 the LQCD results are in good agreement with the experimental average.

There is an important observation about the dipole parameterization here.
There is a one-to-one correspondence between the squared axial radius
 and the dipole axial mass parameter:
\begin{align}
 r_{A, {\rm dipole}}^2 = 12/m_A^2.
\end{align}
If both the $g_A$ and $r_A^2$ are known,
 then the $Q^2$ dependence of dipole parameterization is completely fixed.
If the dipole parameterization for the axial form factor is assumed to be correct,
 then the agreement with the squared axial radius in Fig.~\ref{fig:ra2}
 is in contradiction with the high $Q^2$ discrepancy in Fig.~\ref{fig:gaq2-summary}.
Either both would have to agree,
 meaning the LQCD curves should be as low as the deuterium $z$ expansion results,
 or both results would have to disagree, resulting in an axial radius
 as low as $r_{A}^2 \approx 0.25~{\rm fm}^2$.
This is a handwaving argument about the inconsistency of the LQCD form factors
 with the dipole parameterization;
 the LQCD results could be fit to a dipole parameterization,
 but such a fit would result in a poor fit quality.
A small squared axial radius would be obtained,
 resulting from the fit trying to compensate for the high form factor values at large $Q^2$.

\begin{figure}[htb!]
\centering
\includegraphics[width=0.7\textwidth]{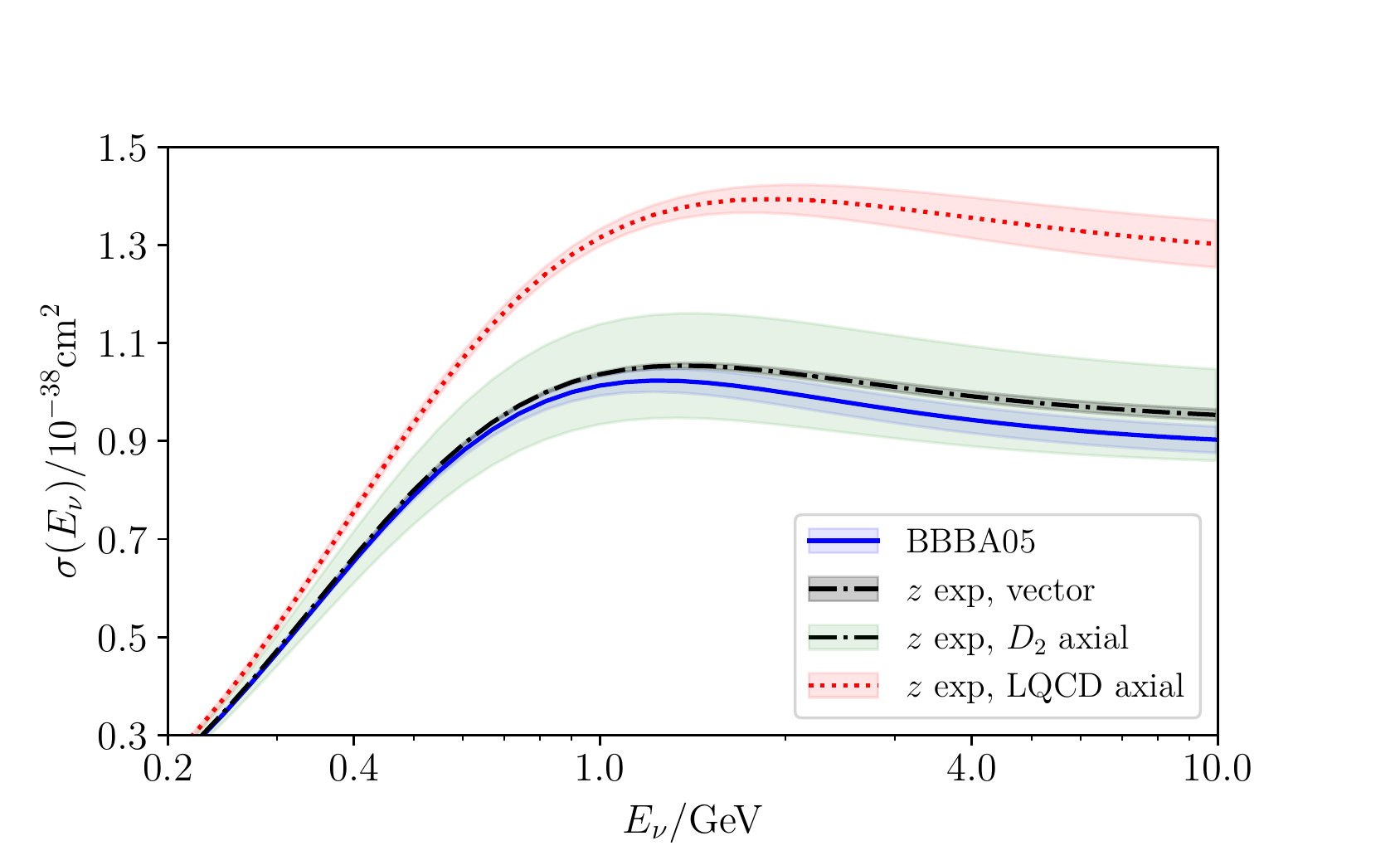}
\caption{
A comparison of results for the neutrino energy dependence
 of the CCQE cross section for a muon neutrino interaction with a neutron target,
 computed using various form factor parameterizations.
Both the black and green bands (dot-dashed line) are computed with
 the vector form factors from Ref.~\cite{Borah:2020gte}
 and the axial form factor from Ref.~\cite{Meyer:2016oeg}.
The green band uses the uncertainty only from the axial form factor,
 while the black band takes the uncertainty only from the vector form factors.
The red band (dotted line) is the same as the green band,
 except replaces the axial form factor central value and uncertainty
 with the parameterization from Ref.~\cite{Meyer:2021vfq}.
The blue band (solid line) is the same as the black band,
 except replaces the vector form factor central values and uncertainties
 with the parameterization from Ref.~\cite{Bradford:2006yz}.
Figure reproduced from Ref.~\cite{Meyer:2022mix}.
\label{fig:xsec-comparison}
}
\end{figure}

The axial form factor results shown in Fig.~\ref{fig:gaq2-summary}
 can also be integrated into a nucleon quasielastic cross section
 for a muon neutrino interaction with a free neutron.
This total CCQE cross section is plotted in Fig.~\ref{fig:xsec-comparison},
 comparing the deuterium scattering parameterization
 of Ref.~\cite{Meyer:2016oeg} (dot-dash, green band)
 against a proxy LQCD result from Ref.~\cite{Meyer:2021vfq} (dotted, red band).
The results are shocking:
 the integration over $Q^2$ enhances the tension between the
 deuterium scattering and the LQCD results,
 suggesting an approximate 30\% enhancement of the CCQE cross section is needed.
Although this seems like a drastic shift,
 there is some experimental evidence that such an enhancement could be a real effect.
Two recent Monte Carlo tunes to neutrino cross section data,
 from MicroBooNE~\cite{MicroBooNE:2021ccs} and GENIE~\cite{GENIE:2022qrc},
 note that the CCQE cross section needs to be enhanced by at least
 20\% to produce a reasonable description of the experimental data.

The two remaining bands in Fig.~\ref{fig:xsec-comparison},
 compare two different vector form factor parameterizations.
One is the BBBA05 parameterization~\cite{Bradford:2006yz} (solid, blue),
 which is commonly used in neutrino event generators,
 compared against a new extraction using the $z$ expansion
 from Ref.~\cite{Borah:2020gte} (dot-dashed, black).
For both of these curves, the uncertainty band is computed for only
 from the vector form factors.
These two parameterizations have a slight $1$--$1.5\sigma$ tension
 with each other, originating from a disagreement
 for the proton magnetic form factor.
However, the size of the (red) uncertainty band for the LQCD axial form factor calculation
 is the same size as this tension.
This indicates that cross section uncertainties will soon be limited
 not by the axial form factor precision, but instead by the vector form factor tension.
LQCD could resolve the tension with a precise calculation of the slope of the
 isovector magnetic form factor.
Additional, complementary constraints from calculations of
 the vector isoscalar form factors, which are only constrained to the 20--50\% level,
 can also make important contributions to better fix these vector form factors.

\subsection{Experimental Implications}
\label{sec:implications}

Although the nucleon CCQE cross section is an important benchmark,
 oscillation experiments need predictions for neutrino event rates
 with nuclear targets.
Monte Carlo event generators, such as GENIE~\cite{Andreopoulos:2009rq},
 are needed to handle the nuclear modeling and final state interactions.
These generators include a variety of initial event topologies,
 including but not limited to CCQE.

\begin{figure}[b!ht]
\centering
\includegraphics[width=0.4\textwidth]{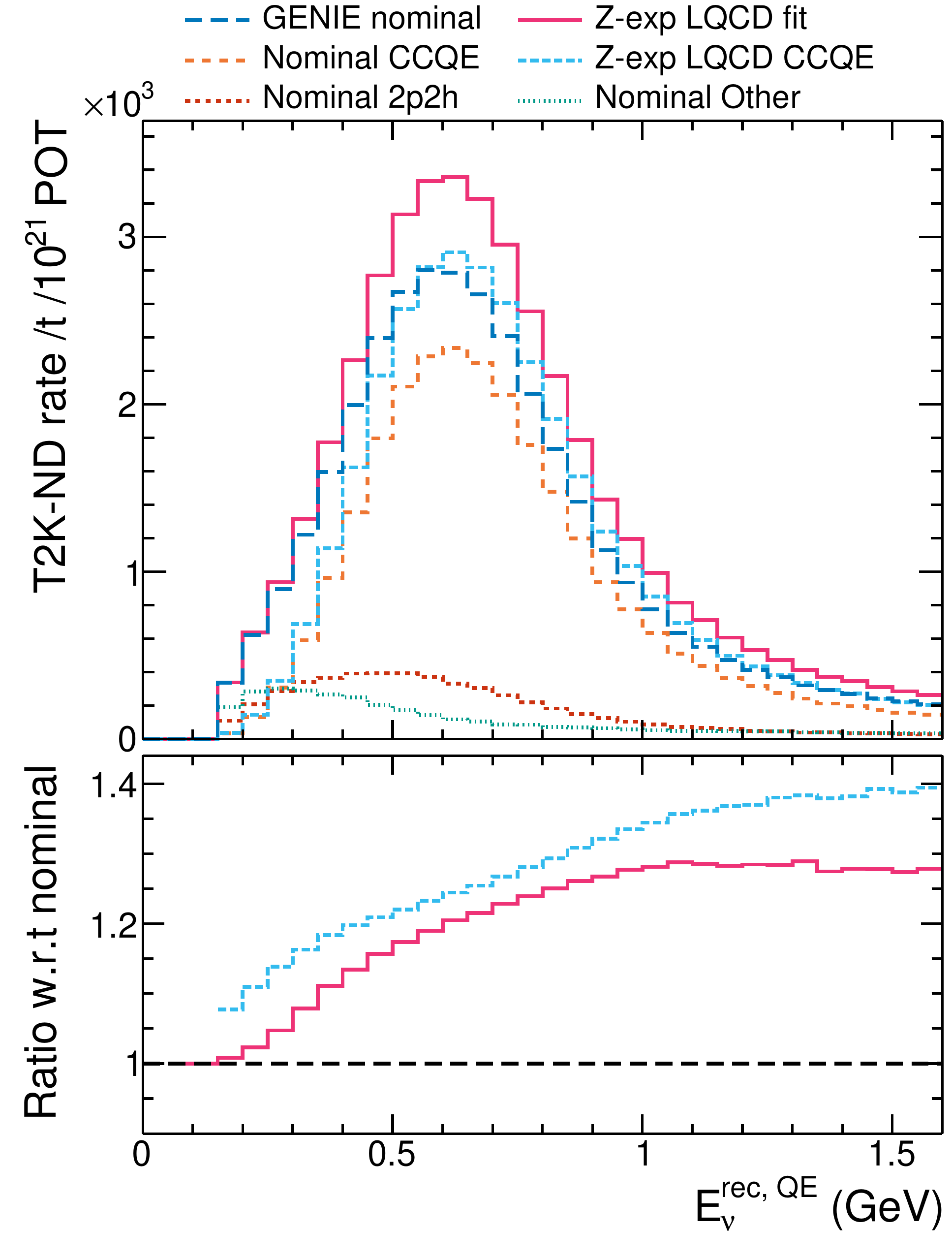}
\includegraphics[width=0.4\textwidth]{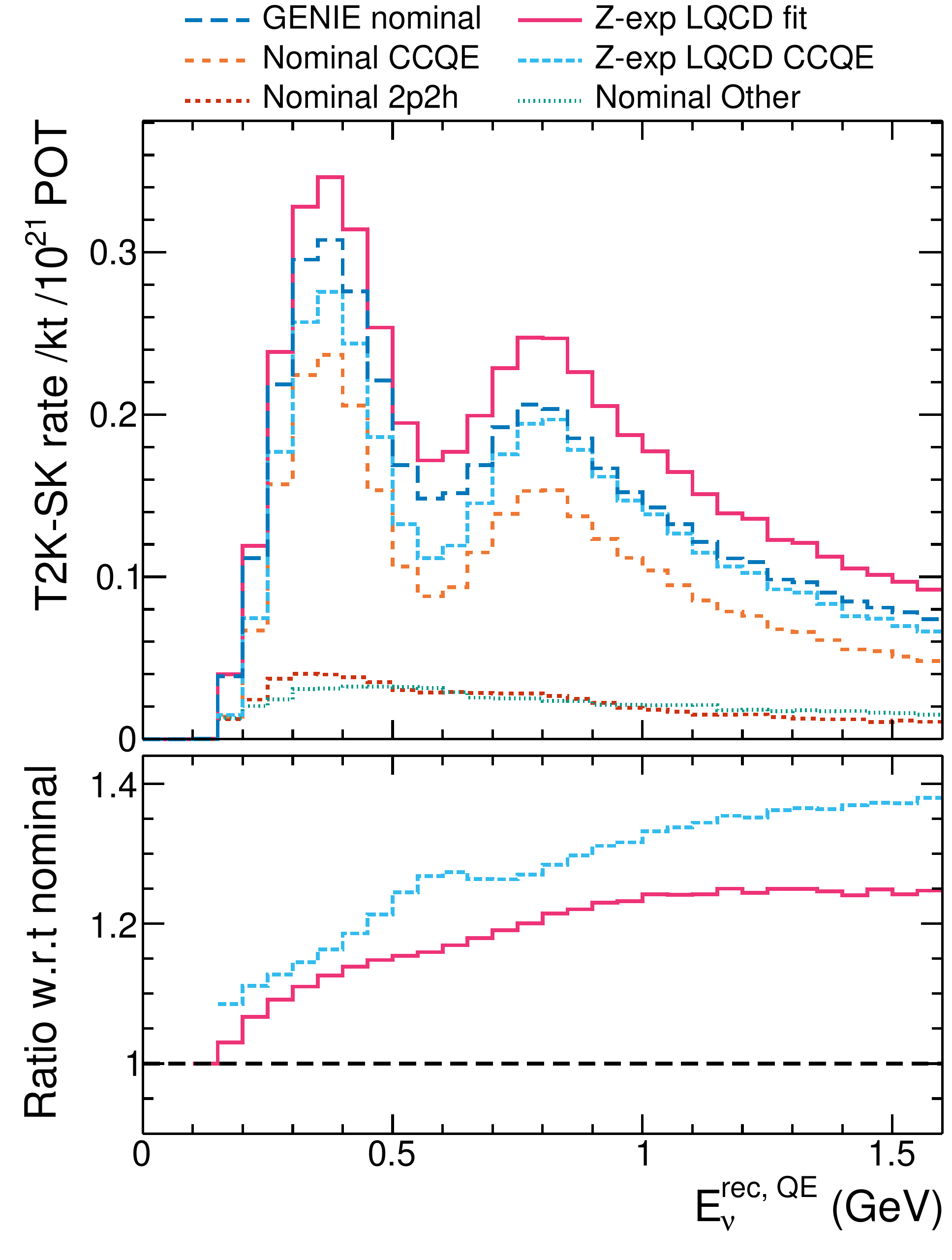}
\caption{
The effect of the axial form factor parameterization
 on the T2K near and far detector event rates.
The horizontal axis is the reconstructed neutrino energy
 obtained from charged lepton assuming quasielastic scattering.
The vertical axis is the neutrino event rate.
The left panels are near detector predictions,
 and the right panels are far detector predictions.
In all of the panels, the dashed dark blue curve is the
 nominal GENIE prediction for the total event rate.
The nominal prediction is broken down into CCQE, 2p2h, and other contributions.
In the top panels, which show the properly normalized neutrino event rates,
 the ``$z$-exp LQCD fit'' (``$z$-exp LQCD CCQE'')
 is the same as the GENIE nominal (nominal CCQE),
 except with the axial form factor taken from Ref.~\cite{Meyer:2021vfq}.
In the bottom panels, the solid and dashed curves give the ratio
 of the $z$-exp LQCD event rates divided by the nominal event rates
 for the total and CCQE contributions, respectively.
Figure reproduced from Ref.~\cite{Meyer:2022mix}.
\label{fig:t2k-comp}
}
\end{figure}

\begin{figure}[htb!]
\centering
\includegraphics[width=0.4\textwidth]{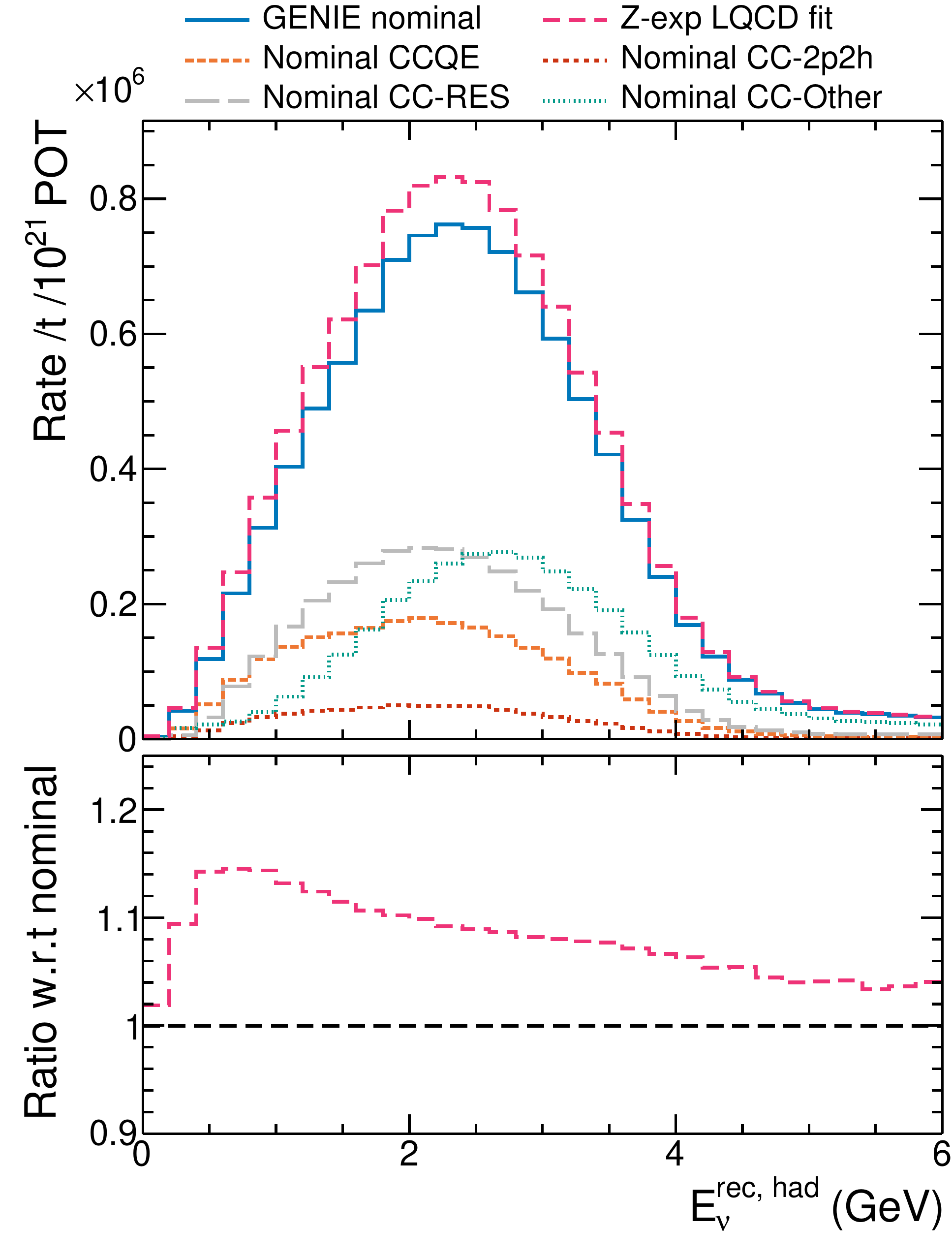}
\includegraphics[width=0.4\textwidth]{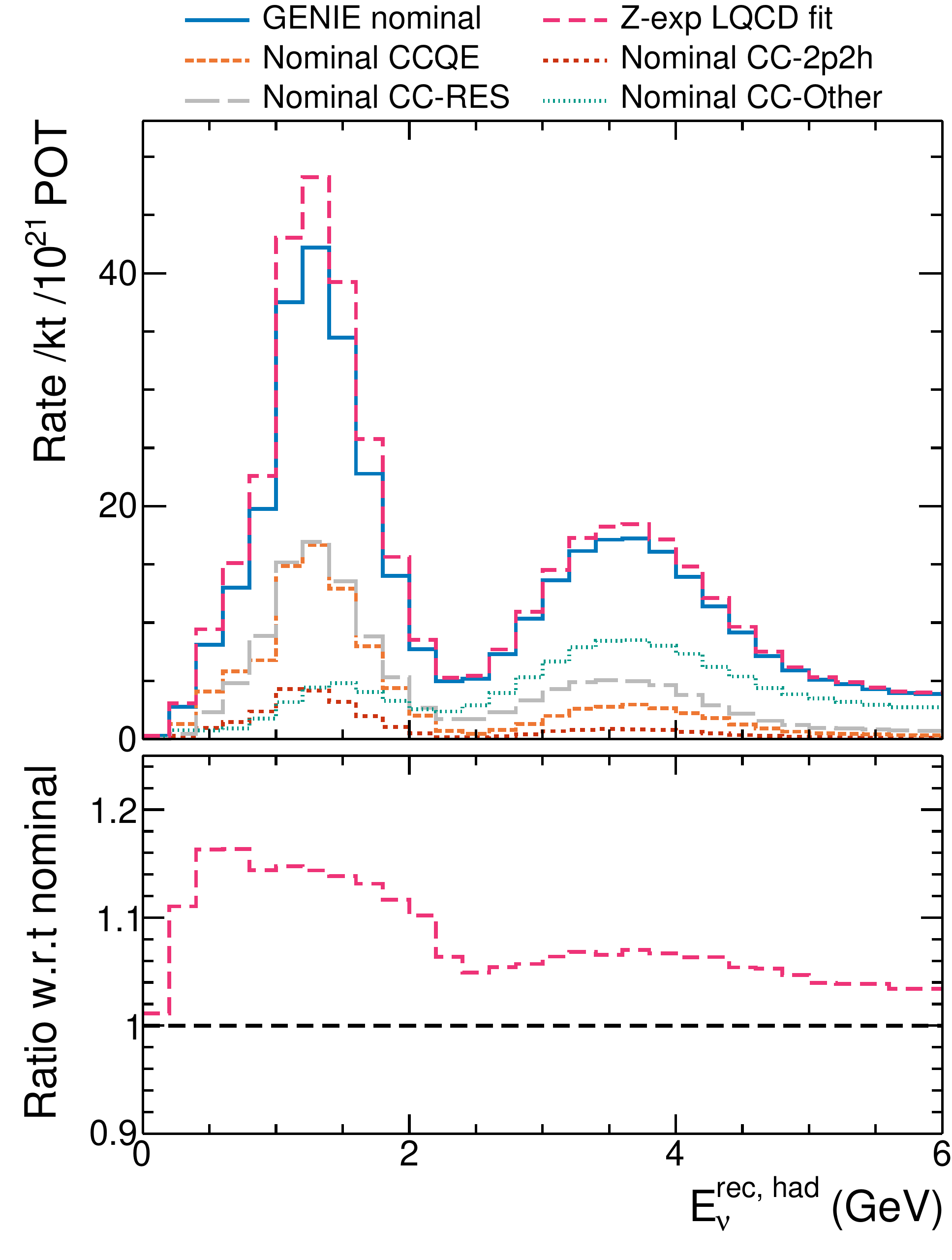}
\caption{
Similar to Fig.~\ref{fig:t2k-comp}, but instead showing the event rates for DUNE.
Like in Fig.~\ref{fig:t2k-comp}, the top panel shows properly normalized event rates
 and the bottom panel the ratios of event rates, and the left and right panels
 correspond to near and far detector predictions, respectively.
The horizontal axis is the neutrino energy reconstructed from hadronic visible energy.
The GENIE nominal ($z$-exp LQCD fit) is the solid blue (dashed magenta)
 line in the top panels.
The bottom panel only shows the ratio of $z$-exp LQCD fit over the GENIE nominal.
Figure reproduced from Ref.~\cite{Meyer:2022mix}.
\label{fig:dune-comp}
}
\end{figure}

Figs.~\ref{fig:t2k-comp}~and~\ref{fig:dune-comp} show the effect of switching the
 nominal CCQE axial form factor from GENIE v3, with the 10a\_02\_11a tune,
 to the axial form factor obtained from LQCD data in Ref.~\cite{Meyer:2021vfq}.
Fig.~\ref{fig:t2k-comp} shows the effect on the T2K experiment, with a water target,
 and Fig.~\ref{fig:dune-comp} for DUNE, a liquid argon target.
In both figures, the top panels show the properly normalized event rates,
 and the bottom panels are ratios of event rates.
The left panels are for the near detector, before the neutrinos have been able
 to oscillate appreciably, and the right panels show the event rates after oscillation.
In both cases, the ``$z$-exp LQCD'' results are obtained by replacing
 only the axial form factor and keeping all other inputs fixed.

For T2K, the peak beam energy is lower than for DUNE.
The resulting event rates are predominantly CCQE, 
 with some contribution from resonance scattering.
It is therefore reasonable to make the assumption of quasielastic
 scattering and compute a reconstructed neutrino energy
 from the outgoing lepton kinematics assuming a struck neutron at rest,
\begin{align}
 E_{\nu}^{\rm rec, QE}( p_\ell, \theta_\ell)
 = \frac{2m_f \sqrt{p_\ell^2 +m_\ell^2} -m_\ell^2 +m_i^2 -m_f^2}{
   2 \big( m_f -\sqrt{p_\ell^2+m_\ell^2} +p_\ell \cos\theta_\ell \big)} .
\end{align}
This reconstruction is applied to events to the ``CC0$\pi$'' event sample,
 which contains an outgoing muon, no mesons, and any number of outgoing nucleons.

In Fig.~\ref{fig:t2k-comp}, comparing the ``GENIE nominal'' (dark blue, long dashes)
 to the ``Nominal CCQE'' (orange, short dashes) shows how much of the event
 sample comes from the CCQE events that are well described by
 the quasielastic assumption.
After replacing the axial form factor,
 the ``GENIE nominal'' becomes the ``$z$-exp LQCD fit'' (magenta, solid)
 and the ``Nominal CCQE'' becomes the ``$z$-exp LQCD CCQE'' (light blue, short dashes).
The ratio of these two choices,
 both ``GENIE nominal'' over ``$z$-exp LQCD fit'' (magenta, solid)
 and ``Nominal CCQE'' over ``$z$-exp LQCD CCQE'' (light blue, short dashes),
 are plotted in the bottom panel.
The bottom panel shows that the replacement of the axial form factor 
 enhances the CCQE contribution by as much as 35\%,
 while the total event rate is enhanced up to 20\%.

For DUNE, the peak beam energy is significantly higher
 and so many interaction processes play a role in the event rates.
It is more advantageous to instead consider an inclusive charge current, ``CC-inclusive,''
 sample and to reconstruct the neutrino energy from the visible hadronic energy,
\begin{align}
 E_{\nu}^{\rm rec, had}
 = E_\ell
 + \sum_{\rm proton} E_{\rm KE}
 + \sum_{\pi^{\pm}, \pi^0, \gamma} E_{\rm total}.
\end{align}
This sums the total lepton energy,
 the proton kinetic energy,
 and the total energy from pions and photons,
 which is an approximation of the expected measurable energy
 from a neutrino event.

In Fig.~\ref{fig:dune-comp}, the total ``GENIE nominal'' (dark blue, solid)
 becomes the ``$z$-exp LQCD fit'' (magenta, dashed) after the replacement
 of the form factor.
The ratio of these two choices,
 ``GENIE nominal'' over ``$z$-exp LQCD fit'' (magenta, dashed),
 is again plotted in the bottom panel.
Like Fig.~\ref{fig:t2k-comp}, the bottom panel again shows that the
 replacement of the axial form factor enhances the total event rate
 by up to 15\%.
However, unlike Fig.~\ref{fig:t2k-comp},
 the enhancement \emph{decreases} over most of the range
 with increasing reconstructed energy.
There is also a significant feature in the far detector
 event rate prediction centered directly over the oscillation dip.

The more troublesome features of this enhancement
 are its dependence on the reconstructed neutrino energy
 and the differences between the near and far detectors.
The Monte Carlo parameters are typically tuned to the measured event
 rates in the near detector.
If the assumed model is not able to accommodate the changes
 to the CCQE form factors, then tuning to the near detector data
 will introduce a bias for the far detector prediction.
This bias could adversely affect oscillation measurements,
 causing an apparent shift of the location of the oscillation dip
 of artificially increasing or decreasing its depth.

This study is representative of some of the issues that will be faced
 by neutrino oscillation experiments, but it is not a complete picture.
The enhancement of the event rate due to the slow falloff with $Q^2$
 of the form factor will be compensated by a corresponding reduction
 of another event topology, for instance the 2p2h contribution
 from correlated nucleons.
This is not a straightforward task since certain event topologies
 more strongly affect certain kinematic regions than others.
Both the T2K and DUNE event rate predictions will suffer from these
 potential biases, and so care must be taken to ensure that the oscillation
 model is sufficiently flexible to accommodate such changes.

\section{Future Prospects}
\label{sec:future}

The nucleon axial form factor is just the beginning of full programs
 of LQCD calculations that can produce useful results for
 the long baseline neutrino oscillation experimental program.
Mature computations with complete uncertainty budgets,
 physical pion masses, explicit $N\pi$ operators, and demonstrated robust control
 over excited state contamination are already available or will soon be available.
Within the next few years, the emerging consensus within the community
 will most likely be demonstrated and meaningful averages of nucleon form factors
 from LQCD can be created and included in neutrino Monte Carlo event generators.
Precise constraints on the slope of the isovector magnetic form factor
 and better than 20\% constraints on either of the isoscalar form factors
 can have already weigh in on mild but apparent tensions
 the will soon be limiting factors on the precision of neutrino-nucleon cross sections.

Beyond quasielastic scattering, LQCD will also have important implications
 for producing missing inputs to predictions of event rates
 for one or two pion production in both HyperK and DUNE.
Form factors for nucleon transitions to the $\Delta$ and other baryonic resonances
 are difficult to constrain experimentally even with electron-nucleon scattering,
 and even more so when mediated by a weak current.
Constraints from spin-$\frac{3}{2}$ resonances are especially important,
 where interference from axial amplitudes with each other
 or the vector amplitudes are poorly constrained by experiments
 and have 100\% uncertainties~\cite{Lalakulich:2006sw}.
LQCD will provide amplitudes for these transitions
 decomposed into definite spin and isospin interaction channels.
Since these calculations access the matrix elements rather than the squared amplitude,
 they can provide valuable constraints for disentangling interference effects.

Between resonant pion production and deep inelastic scattering
 is the ``shallow inelastic scattering'' region,
 which is inconveniently situated between scales.
This region of parameter space has energies and momentum transfers
 too high to be represented by the response of a few resonances
 or creation of one or two pions,
 and yet too low to be computed with perturbative QCD.
In this regime, there are essentially no experimental constraints
 for hadronic amplitudes~\cite{NuSTEC:2019lqd}.
However, as much as a third of neutrino interaction events
 for DUNE will fall in this kinematic regime.
LQCD calculations could potentially provide inputs where
 with four-point functions will constrain the
 hadronic tensor amplitude.

With advances in computational techniques and increased availability of computing power,
 additional strategies for handling excited state contamination are becoming feasible.
Adaption of all-to-all strategies could allow for more comprehensive analyses,
 for instance including momentum projections at both source and sink rather than just one.
This would overconstrain the determination of form factors, providing
 additional ability to disentangle excited state contributions and other systematics.
However, attention should be paid to the covariance matrix conditioning:
 for current analyses with upwards of 30 choices of insertion momenta,
 it is already borderline to have only $O(10^3)$ measurements.
Better fitting techniques that are not as sensitive to correlations must be used,
 or more measurements are needed to better condition the problem.

\section{Concluding Remarks}
\label{sec:conclusions}

Neutrino event rate predictions from elementary targets have a long history.
Deuterium bubble chamber scattering experiments from 40 years ago
 have traditionally been used to claim 1\% uncertainties on nucleon form factors.
The onus of resolving theory--experiment discrepancies in neutrino scattering
 was consequently placed onto nuclear theory.
Application of more modern techniques have demonstrated that
 these problems could be more than just from nuclear modeling.
Model independent determinations of uncertainties on nucleon form factors 
 from elementary target sources are a factor of 10 larger than previously though,
 shifting the blame of discrepancies to some unknown combination
 of nucleon and/or nuclear origin.
Still, issues with the deuterium scattering data remain,
 especially in regard to the corrections from the spectator proton.
Stronger constraints of nucleon-level amplitudes for weak interactions are needed,
 yet no viable experimental solution is yet forthcoming.

As a complement to new experimental solutions on elementary targets,
 LQCD offers a way to probe nucleon physics from first principles computations.
Calculations of $g_A$,
 a key benchmark quantity for nucleon computations in LQCD,
 have traditionally been underestimated compared to experiment.
However, recent advances have demonstrated the importance of nucleon-pion
 excited states for understanding this discrepancy.
This success has extended to the momentum transfer dependence of the form factor,
 which pass checks using PCAC and PPD relations.
Now, computations with complete uncertainty budgets are becoming available
 and so far no significant tensions have persisted between LQCD calculations.

For the purposes of improving upon elementary target constraints of nucleon physics,
 LQCD computations are already proving to have real impact for
 event rate predictions in long baseline neutrino oscillation experiments.
Mounting evidence from LQCD indicates that the large $Q^2$ region of the axial
 form factor has been significantly underestimated,
 well outside of even the 10\% uncertainty from model independent estimates.
This amounts to as much as a 30\% enhancement of the quasielastic cross section.
Despite this apparently large shift, tuning event generators to experimental
 data also suggests a need for a large $\sim$20\% enhancement
 of the quasielastic cross section.
This enhancement produces neutrino energy dependent shifts to the
 event rates for both T2K and DUNE, risking the introduction of bias
 for event generator models that are not sufficiently flexible to accommodate the changes.

It is clear that LQCD results will remain important for understanding
 cross sections and event rates in long baseline neutrino oscillation experiments
 for decades to come.
Some of the first works with nucleon matrix elements are already proving their worth,
 overturning decades of accepted wisdom about nucleon form factors.
As these computations become better understood,
 progress will continue on new and more sophisticated computations
 for higher energy processes involving multiparticle states.
These computations will provide valuable information about matrix elements
 and their interference that are impractical or inaccessible to experimental constraints.
LQCD is therefore a powerful tool that will enable theorists to make important contributions
 in support of experimental efforts.
Within the next several years, the LQCD community will take many important steps
 that will bolster the physics programs of flagship neutrino oscillation experiments.

\section{Acknowledgements}

I wish to thank
 Rajan Gupta,
 Andreas Kronfeld,
 Yin Lin,
 Keh-Fei Liu,
 Andr\'e Walker-Loud,
 and my other CalLat, Fermilab Lattice, and N3AS collaborators for useful discussions.
I wish to thank
 Constantia Alexandrou,
 Lorenzo Barca,
 Rajan Gupta,
 Keh-Fei Liu,
 Shigemi Ohta,
 and
 Shoichi Sasaki
 for providing me with preliminary results
 and for discussing those results with me.
This work was supported by the Department of Energy, Office of Nuclear Physics,
 under Contract No. DE-SC00046548.


\begin{thebibliography}{99}

\bibitem{DUNE:2020ypp}
B.~Abi \textit{et al.} [DUNE],
[arXiv:2002.03005 [hep-ex]].

\bibitem{Hyper-Kamiokande:2018ofw}
K.~Abe \textit{et al.} [Hyper-Kamiokande],
[arXiv:1805.04163 [physics.ins-det]].

\bibitem{Bodek:2007ym}
A.~Bodek, S.~Avvakumov, R.~Bradford and H.~S.~Budd,
Eur. Phys. J. C \textbf{53}, 349-354 (2008)
doi:10.1140/epjc/s10052-007-0491-4
[arXiv:0708.1946 [hep-ex]].

\bibitem{Bhattacharya:2011ah}
B.~Bhattacharya, R.~J.~Hill and G.~Paz,
Phys. Rev. D \textbf{84}, 073006 (2011)
doi:10.1103/PhysRevD.84.073006
[arXiv:1108.0423 [hep-ph]].

\bibitem{Bernard:2001rs}
V.~Bernard, L.~Elouadrhiri and U.~G.~Meissner,
J. Phys. G \textbf{28}, R1-R35 (2002)
doi:10.1088/0954-3899/28/1/201
[arXiv:hep-ph/0107088 [hep-ph]].

\bibitem{Baker:1981su}
N.~J.~Baker, A.~M.~Cnops, P.~L.~Connolly, S.~A.~Kahn, H.~G.~Kirk, M.~J.~Murtagh, R.~B.~Palmer, N.~P.~Samios and M.~Tanaka,
Phys. Rev. D \textbf{23}, 2499-2505 (1981)
doi:10.1103/PhysRevD.23.2499

\bibitem{Miller:1982qi}
K.~L.~Miller, S.~J.~Barish, A.~Engler, R.~W.~Kraemer, B.~J.~Stacey, M.~Derrick, E.~Fernandez, L.~Hyman, G.~Levman and D.~Koetke, \textit{et al.}
Phys. Rev. D \textbf{26}, 537-542 (1982)
doi:10.1103/PhysRevD.26.537

\bibitem{Kitagaki:1983px}
T.~Kitagaki, S.~Tanaka, H.~Yuta, K.~Abe, K.~Hasegawa, A.~Yamaguchi, K.~Tamai, T.~Hayashino, Y.~Otani and H.~Hayano, \textit{et al.}
Phys. Rev. D \textbf{28}, 436-442 (1983)
doi:10.1103/PhysRevD.28.436

\bibitem{Meyer:2016oeg}
A.~S.~Meyer, M.~Betancourt, R.~Gran and R.~J.~Hill,
Phys. Rev. D \textbf{93}, no.11, 113015 (2016)
doi:10.1103/PhysRevD.93.113015
[arXiv:1603.03048 [hep-ph]].

\bibitem{FlavourLatticeAveragingGroup:2019iem}
S.~Aoki \textit{et al.} [Flavour Lattice Averaging Group],
Eur. Phys. J. C \textbf{80}, no.2, 113 (2020)
doi:10.1140/epjc/s10052-019-7354-7
[arXiv:1902.08191 [hep-lat]].

\bibitem{Kronfeld:2019nfb}
A.~S.~Kronfeld \textit{et al.} [USQCD],
Eur. Phys. J. A \textbf{55}, no.11, 196 (2019)
doi:10.1140/epja/i2019-12916-x
[arXiv:1904.09931 [hep-lat]].

\bibitem{Capitani:2012gj}
S.~Capitani, M.~Della Morte, G.~von Hippel, B.~Jager, A.~Juttner, B.~Knippschild, H.~B.~Meyer and H.~Wittig,
Phys. Rev. D \textbf{86}, 074502 (2012)
doi:10.1103/PhysRevD.86.074502
[arXiv:1205.0180 [hep-lat]].

\bibitem{Bar:2016uoj}
O.~B\"ar,
Phys. Rev. D \textbf{94}, no.5, 054505 (2016)
doi:10.1103/PhysRevD.94.054505
[arXiv:1606.09385 [hep-lat]].

\bibitem{Bar:2017kxh}
O.~Bar,
Int. J. Mod. Phys. A \textbf{32}, no.15, 1730011 (2017)
doi:10.1142/S0217751X17300113
[arXiv:1705.02806 [hep-lat]].

\bibitem{He:2021yvm}
J.~He, D.~A.~Brantley, C.~C.~Chang, I.~Chernyshev, E.~Berkowitz, D.~Howarth, C.~K\"orber, A.~S.~Meyer, H.~Monge-Camacho and E.~Rinaldi, \textit{et al.}
Phys. Rev. C \textbf{105}, no.6, 065203 (2022)
doi:10.1103/PhysRevC.105.065203
[arXiv:2104.05226 [hep-lat]].

\bibitem{Jang:2019vkm}
Y.~C.~Jang, R.~Gupta, B.~Yoon and T.~Bhattacharya,
Phys. Rev. Lett. \textbf{124}, no.7, 072002 (2020)
doi:10.1103/PhysRevLett.124.072002
[arXiv:1905.06470 [hep-lat]].

\bibitem{Bar:2018xyi}
O.~Bar,
Phys. Rev. D \textbf{99}, no.5, 054506 (2019)
doi:10.1103/PhysRevD.99.054506
[arXiv:1812.09191 [hep-lat]].

\bibitem{Bar:2019gfx}
O.~Bar,
Phys. Rev. D \textbf{100}, no.5, 054507 (2019)
doi:10.1103/PhysRevD.100.054507
[arXiv:1906.03652 [hep-lat]].

\bibitem{Bar:2019igf}
O.~Bar,
Phys. Rev. D \textbf{101}, no.3, 034515 (2020)
doi:10.1103/PhysRevD.101.034515
[arXiv:1912.05873 [hep-lat]].

\bibitem{Wilson:2015dqa}
D.~J.~Wilson, R.~A.~Briceno, J.~J.~Dudek, R.~G.~Edwards and C.~E.~Thomas,
Phys. Rev. D \textbf{92}, no.9, 094502 (2015)
doi:10.1103/PhysRevD.92.094502
[arXiv:1507.02599 [hep-ph]].

\bibitem{Meyer:2022mix}
A.~S.~Meyer, A.~Walker-Loud and C.~Wilkinson,
doi:10.1146/annurev-nucl-010622-120608
[arXiv:2201.01839 [hep-lat]].

\bibitem{Alexandrou:2020okk}
C.~Alexandrou, S.~Bacchio, M.~Constantinou, P.~Dimopoulos, J.~Finkenrath, K.~Hadjiyiannakou, K.~Jansen, G.~Koutsou, B.~Kostrzewa and T.~Leontiou, \textit{et al.}
Phys. Rev. D \textbf{103}, no.3, 034509 (2021)
doi:10.1103/PhysRevD.103.034509
[arXiv:2011.13342 [hep-lat]].

\bibitem{Park:2021ypf}
S.~Park \textit{et al.} [Nucleon Matrix Elements (NME)],
Phys. Rev. D \textbf{105}, no.5, 054505 (2022)
doi:10.1103/PhysRevD.105.054505
[arXiv:2103.05599 [hep-lat]].

\bibitem{RQCD:2019jai}
G.~S.~Bali \textit{et al.} [RQCD],
JHEP \textbf{05}, 126 (2020)
doi:10.1007/JHEP05(2020)126
[arXiv:1911.13150 [hep-lat]].

\bibitem{Djukanovic:2022wru}
D.~Djukanovic, G.~von Hippel, J.~Koponen, H.~B.~Meyer, K.~Ottnad, T.~Schulz and H.~Wittig,
Phys. Rev. D \textbf{106}, no.7, 074503 (2022)
doi:10.1103/PhysRevD.106.074503
[arXiv:2207.03440 [hep-lat]].

\bibitem{Abramczyk:2019fnf}
M.~Abramczyk, T.~Blum, T.~Izubuchi, C.~Jung, M.~Lin, A.~Lytle, S.~Ohta and E.~Shintani,
Phys. Rev. D \textbf{101}, no.3, 034510 (2020)
doi:10.1103/PhysRevD.101.034510
[arXiv:1911.03524 [hep-lat]].

\bibitem{Tsuji:2022ric}
R.~Tsuji \textit{et al.} [PACS],
Phys. Rev. D \textbf{106}, no.9, 094505 (2022)
doi:10.1103/PhysRevD.106.094505
[arXiv:2207.11914 [hep-lat]].

\bibitem{Tsuji:2021bdp}
R.~Tsuji \textit{et al.} [PACS],
PoS \textbf{LATTICE2021}, 504 (2022)
doi:10.22323/1.396.0504
[arXiv:2112.15276 [hep-lat]].

\bibitem{Shintani:2018ozy}
E.~Shintani, K.~I.~Ishikawa, Y.~Kuramashi, S.~Sasaki and T.~Yamazaki,
Phys. Rev. D \textbf{99}, no.1, 014510 (2019)
[erratum: Phys. Rev. D \textbf{102}, no.1, 019902 (2020)]
doi:10.1103/PhysRevD.99.014510
[arXiv:1811.07292 [hep-lat]].

\bibitem{Ishikawa:2021eut}
K.~I.~Ishikawa \textit{et al.} [PACS],
Phys. Rev. D \textbf{104}, no.7, 074514 (2021)
doi:10.1103/PhysRevD.104.074514
[arXiv:2107.07085 [hep-lat]].

\bibitem{Meyer:2021vfq}
A.~S.~Meyer, E.~Berkowitz, C.~Bouchard, C.~C.~Chang, M.~A.~Clark, B.~H\"orz, D.~Howarth, C.~K\"orber, H.~Monge-Camacho and A.~Nicholson, \textit{et al.}
PoS \textbf{LATTICE2021}, 081 (2022)
doi:10.22323/1.396.0081
[arXiv:2111.06333 [hep-lat]].

\bibitem{Hill:2017wgb}
R.~J.~Hill, P.~Kammel, W.~J.~Marciano and A.~Sirlin,
Rept. Prog. Phys. \textbf{81}, no.9, 096301 (2018)
doi:10.1088/1361-6633/aac190
[arXiv:1708.08462 [hep-ph]].

\bibitem{Borah:2020gte}
K.~Borah, R.~J.~Hill, G.~Lee and O.~Tomalak,
Phys. Rev. D \textbf{102}, no.7, 074012 (2020)
doi:10.1103/PhysRevD.102.074012
[arXiv:2003.13640 [hep-ph]].

\bibitem{Bradford:2006yz}
R.~Bradford, A.~Bodek, H.~S.~Budd and J.~Arrington,
Nucl. Phys. B Proc. Suppl. \textbf{159}, 127-132 (2006)
doi:10.1016/j.nuclphysbps.2006.08.028
[arXiv:hep-ex/0602017 [hep-ex]].

\bibitem{MicroBooNE:2021ccs}
P.~Abratenko \textit{et al.} [MicroBooNE],
Phys. Rev. D \textbf{105}, no.7, 072001 (2022)
doi:10.1103/PhysRevD.105.072001
[arXiv:2110.14028 [hep-ex]].

\bibitem{GENIE:2022qrc}
J.~Tena-Vidal \textit{et al.} [GENIE],
Phys. Rev. D \textbf{106}, no.11, 112001 (2022)
doi:10.1103/PhysRevD.106.112001
[arXiv:2206.11050 [hep-ph]].

\bibitem{Andreopoulos:2009rq}
C.~Andreopoulos, A.~Bell, D.~Bhattacharya, F.~Cavanna, J.~Dobson, S.~Dytman, H.~Gallagher, P.~Guzowski, R.~Hatcher and P.~Kehayias, \textit{et al.}
Nucl. Instrum. Meth. A \textbf{614}, 87-104 (2010)
doi:10.1016/j.nima.2009.12.009
[arXiv:0905.2517 [hep-ph]].

\bibitem{Lalakulich:2006sw}
O.~Lalakulich, E.~A.~Paschos and G.~Piranishvili,
Phys. Rev. D \textbf{74}, 014009 (2006)
doi:10.1103/PhysRevD.74.014009
[arXiv:hep-ph/0602210 [hep-ph]].

\bibitem{NuSTEC:2019lqd}
C.~Andreopoulos \textit{et al.} [NuSTEC],
[arXiv:1907.13252 [hep-ph]].

\end{thebibliography}
\end{document}